\documentclass[cits]{PoS}\usepackage{amsmath,cite,color}

\title{Open-Flavour Mesons from the Angle of\\Bethe, Dyson, Salpeter, Schwinger, \textit{et al.}}
\ShortTitle{Open-Flavour Mesons from the Angle of Bethe, Dyson, Salpeter, Schwinger, \textit{et al.}}
\author{Thomas Hilger\\Institute for High Energy Physics, Austrian Academy of Sciences, Nikolsdorfergasse 18,\\A-1050 Vienna, Austria\\Institute of Physics, University of Graz, NAWI Graz, A-8010 Graz, Austria\\E-mail: \email{thomas.hilger@uni-graz.at}}
\author{Mar\'ia G\'omez-Rocha\\European Centre for Theoretical Studies in Nuclear Physics and Related Areas, Villa Tambosi, 38123 Villazzano (Trento), Italy\\E-mail: \email{mariagomezrocha@gmail.com}}
\author{Andreas Krassnigg\\Institute of Physics, University of Graz, NAWI Graz, A-8010 Graz, Austria\\E-mail: \email{andreas.krassnigg@uni-graz.at}}
\author{\speaker{Wolfgang Lucha}\\Institute for High Energy Physics, Austrian Academy of Sciences, Nikolsdorfergasse 18,\\A-1050 Vienna, Austria\\E-mail: \email{Wolfgang.Lucha@oeaw.ac.at}} 

\abstract{Recently, we completed a comprehensive investigation of a huge part of the entire meson spectrum by considering both quarkonia and open-flavour mesons by means of a single common framework which unites the homogeneous Bethe--Salpeter equation that describes mesons as quark--antiquark bound states and the Dyson--Schwinger equation that governs the full quark propagator: Adopting two (as a matter of fact, not extremely diverse) models that attempt to grasp all principal aspects of the effective strong interactions entering identically in both these equations, we derived within this unique setup, for all mesons analysed, their masses and leptonic decay constants as well as, for the pseudoscalar ones among these mesons, their in-hadron condensates. Here, as a kind of promotion or teaser, we give but a few examples of the resulting collections of data, laying the~main emphasis on the dependence of our insights on the effective-interaction model underlying all such outcomes.}

\FullConference{XIII Quark Confinement and the Hadron Spectrum --- Confinement2018\\31 July -- 6 August 2018\\Maynooth University, Ireland}

\begin{document}\section{Aim: Unique Poincar\'e-Covariant Study of Quarkonia and Open-Flavour Mesons}So far, Poincar\'e-covariant meson studies do not treat bound states of quark and antiquark~of the same type (quarkonia) and of two unequal types (open-flavour mesons) by a single setup. In~order to connect the two sides of the same coin, we embark on a comprehensive analysis of the whole meson spectrum by applying exactly identical Poincar\'e-covariant descriptions to all possible combinations of quark flavour \cite{GRHKL,WL@}. Specifically, for each meson bound state $(\bar q\,q')$ of momentum $P,$ composed~of antiquark $\bar q$ and quark $q',$ we derive both mass $M_{\bar q\,q'}$ and leptonic decay constant $f_{\bar q\,q'}$ defined, \emph{e.g.},~by$$\langle0|\!:\!\bar q'(0)\,\gamma_\mu\,\gamma_5\,q(0)\!:\!|(\bar q\,q')(P)\rangle={\rm i}\,f_{\bar q\,q'}\,P_\mu\qquad\mbox{for the pseudoscalar mesons}\ .$$

\section{Merger: Quark Dyson--Schwinger Equation and Meson Bethe--Salpeter Equation}\label{a}Trusting in quantum field theory, we adopt for the global study of quark--antiquark bound states the well-established Bethe--Salpeter approach augmented by the quark Dyson--Schwinger equation. The fact that any such equation belongs to an infinite tower of coupled Dyson--Schwinger equations renders the truncation of the tower inevitable. All not available impact (hence dubbed \emph{unobtainium\/}) of thereby skipped relations on the retained ones has to be mimicked by, \emph{e.g.}, sophisticated ansatzes.

Crucial ingredients to any such approach are, for two bound-state constituents discriminated by a subscript $i=1,2,$ their dressed propagators $S_i(p),$ deducible as solutions of the Dyson--Schwinger equation for the corresponding two-point Green function. In rainbow truncation and if Pauli--Villars regularized at a scale $\Lambda,$ the Dyson--Schwinger equation for the dressed quark propagator $S(p)$ reads\begin{equation}S^{-1}(p)=Z_2\,({\rm i}\,\gamma\cdot p+m_{\rm
b})+\frac{4}{3}\,Z_2^2\!\int^\Lambda\frac{{\rm d}^4q}{(2\pi)^4}\,{\mathcal{G}}\!\left((p-q)^2\right)T_{\mu\nu}(p-q)\,\gamma_\mu\,S(q)\,\gamma_\nu\ ,\label{q}\end{equation}involving the quark wave-function renormalization constant $Z_2$, the bare quark mass $m_{\rm
b}$ related by a mass renormalization constant $Z_m$ to the running quark mass $m_{q}(\mu)$ renormalized at a given~scale~$\mu$,
$$m_{\rm b}=Z_m\,m_q(\mu)\ ,$$the transverse projection operator $T_{\mu\nu}(k)$ as a relic of the free gluon propagator in the Landau~gauge,$$T_{\mu\nu}(k)\equiv\delta_{\mu\nu}-\frac{k_\mu\,k_\nu}{k^2}\ ,$$and an effective interaction ${\mathcal{G}}(k^2),$ constructed such as to encompass (the bulk of) the effects of both full gluon propagator and full quark--gluon vertex entering in the quark Dyson--Schwinger~equation. 

The Bethe--Salpeter formalism encodes a two-fermion bound state of relative momentum $p$ and total momentum $P$ by either the Bethe--Salpeter \emph{wave function\/} $\chi(p;P),$ defined as Fourier transform of the matrix element of the time-ordered product of these fermion fields evaluated between vacuum and bound state, or the Bethe--Salpeter \emph{amplitude\/} $\Gamma(p;P),$ differing by the two fermion~propagators:$$\chi(p;P)\equiv S_1(p+\eta\,P)\,\Gamma(p;P)\,S_2(p-(1-\eta)\,P)\ ,\qquad\eta\in[0,1]\ .$$Both manifestations of bound states are solutions of one and the same \emph{homogeneous\/} Bethe--Salpeter equation. In rainbow--ladder truncation, the quark--antiquark Bethe--Salpeter equation is of the form\begin{equation}\Gamma(p;P)=-\frac{4}{3}\,Z_2^2\!\int^\Lambda\frac{{\rm d}^4q}{(2\pi)^4}\,{\mathcal{G}}\!\left((p-q)^2\right)T_{\mu\nu}(p-q)\,\gamma_\mu\,\chi(q;P)\,\gamma_\nu\ .\label{m}\end{equation}Chiral symmetries are properly embedded if the effective coupling ${\mathcal{G}}(k^2)$ is the same as in Eq.~(\ref{q}).

\section{Exemplification: Masses, Leptonic Decay Constants, and In-Hadron Condensates}In modelling the effective interaction in Eqs.~(\ref{q},\ref{m}), we follow two closely related strategies:\begin{itemize}\item Ref.~\cite{EI} tries to capture also the genuine ultraviolet behaviour of the coupling function ${\mathcal{G}}(k^2);$\item Ref.~\cite{EIa} is content with an easier-to-handle simpler behaviour of the coupling function ${\mathcal{G}}(k^2).$\end{itemize}Employing more than just a single model for the effective couplings enables the estimate of, at~least, part of the systematic uncertainties accompanying the findings of the framework sketched in Sect.~\ref{a}.

Meson quark--antiquark bound states carrying total spin $s=0,1$ and orbital angular momentum $\ell=0,1,2,\dots$ can be classified, with respect to their total angular momentum $J,$ parity $P=(-1)^{\ell+1},$ and (only where applicable) charge-conjugation parity $C=(-1)^{\ell+s},$ in terms of the designations \cite{WL@}\begin{align*}&\mbox{ordinary:}&&J^{P\,C}\in\{
0^{++},0^{-+},1^{++},1^{+-},1^{--},2^{++},2^{-+},2^{--},3^{++},3^{+-},3^{--},4^{++},4^{-+},4^{--},\ldots\}\
,\\&\mbox{exotic:}&&J^{P\,C}\in\{ 0^{+-},0^{--},1^{-+},2^{+-},3^{-+},4^{+-},5^{-+},\ldots\}\ .\end{align*}

\begin{figure}[h]\centering\includegraphics[scale=.38191]{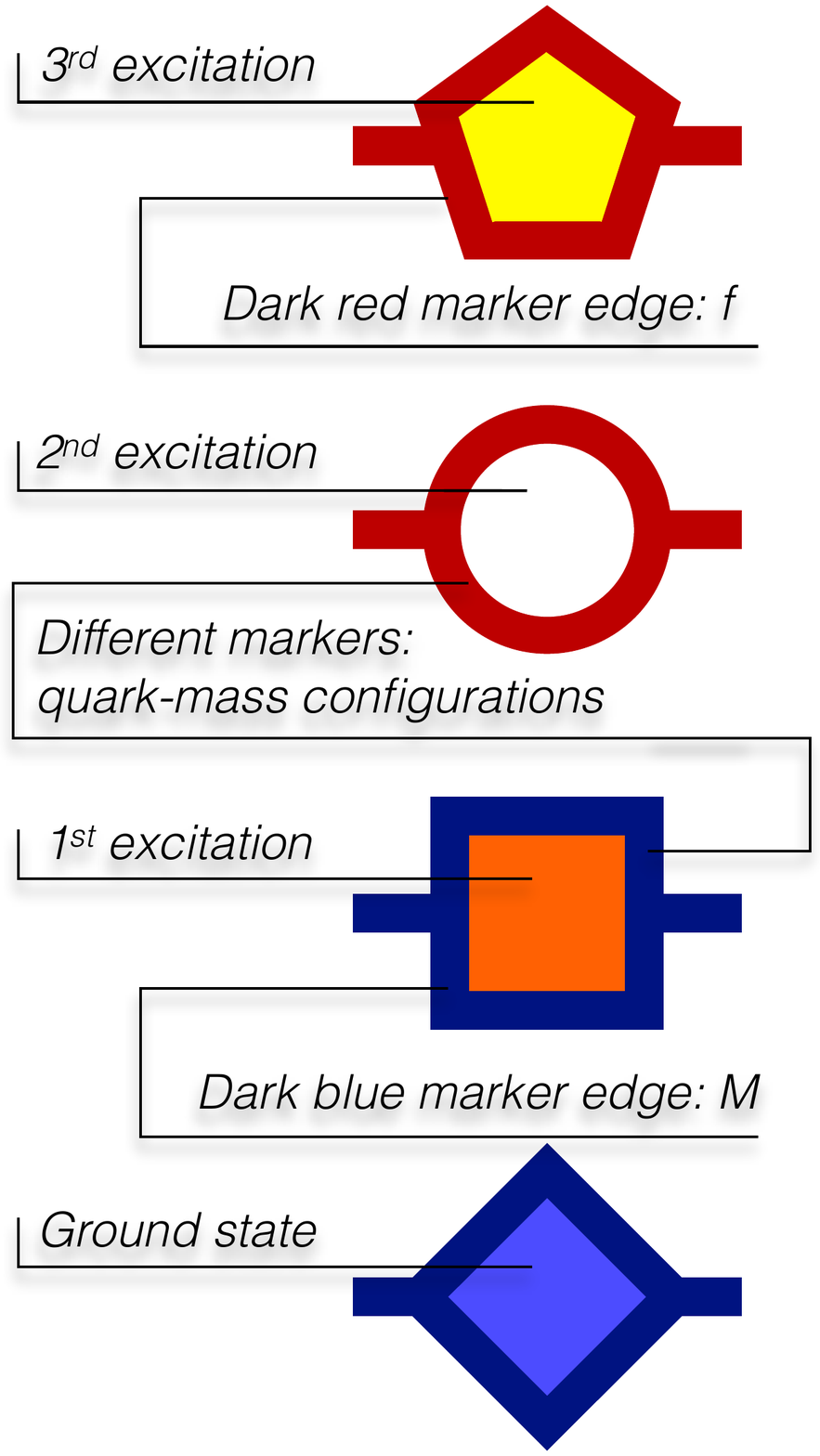}\caption{Legend to Figs.~\protect\ref{qs} through \protect\ref{cc}: in our system, the marker related to some meson bound state~identifies the respective meson features by the colour of its edge (blue for mass and red for leptonic decay constant), the levels of excitation by the colour of its fill (blue, orange, white, yellow, \dots, in increasing order), and the quark combinations in our various evaluations of the quark masses by its shape (diamond, square, circle,~pentagon). Horizontal blue lines delimit the accessible meson-mass ranges originating in the appearance of singularities.}\label{l}\end{figure}\pagebreak

\begin{figure}[h]\begin{center}\includegraphics[scale=.59784]{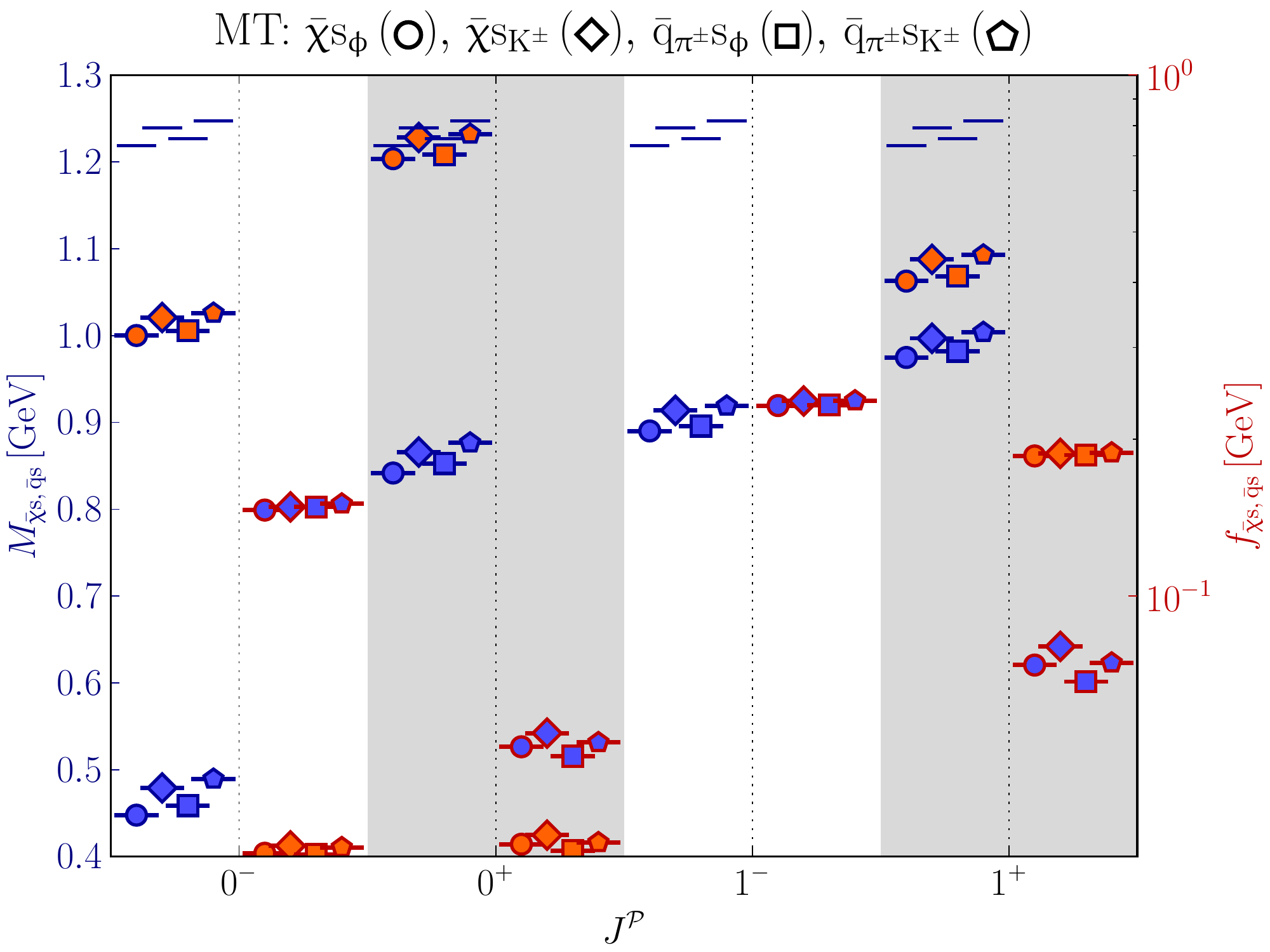} \includegraphics[scale=.59784]{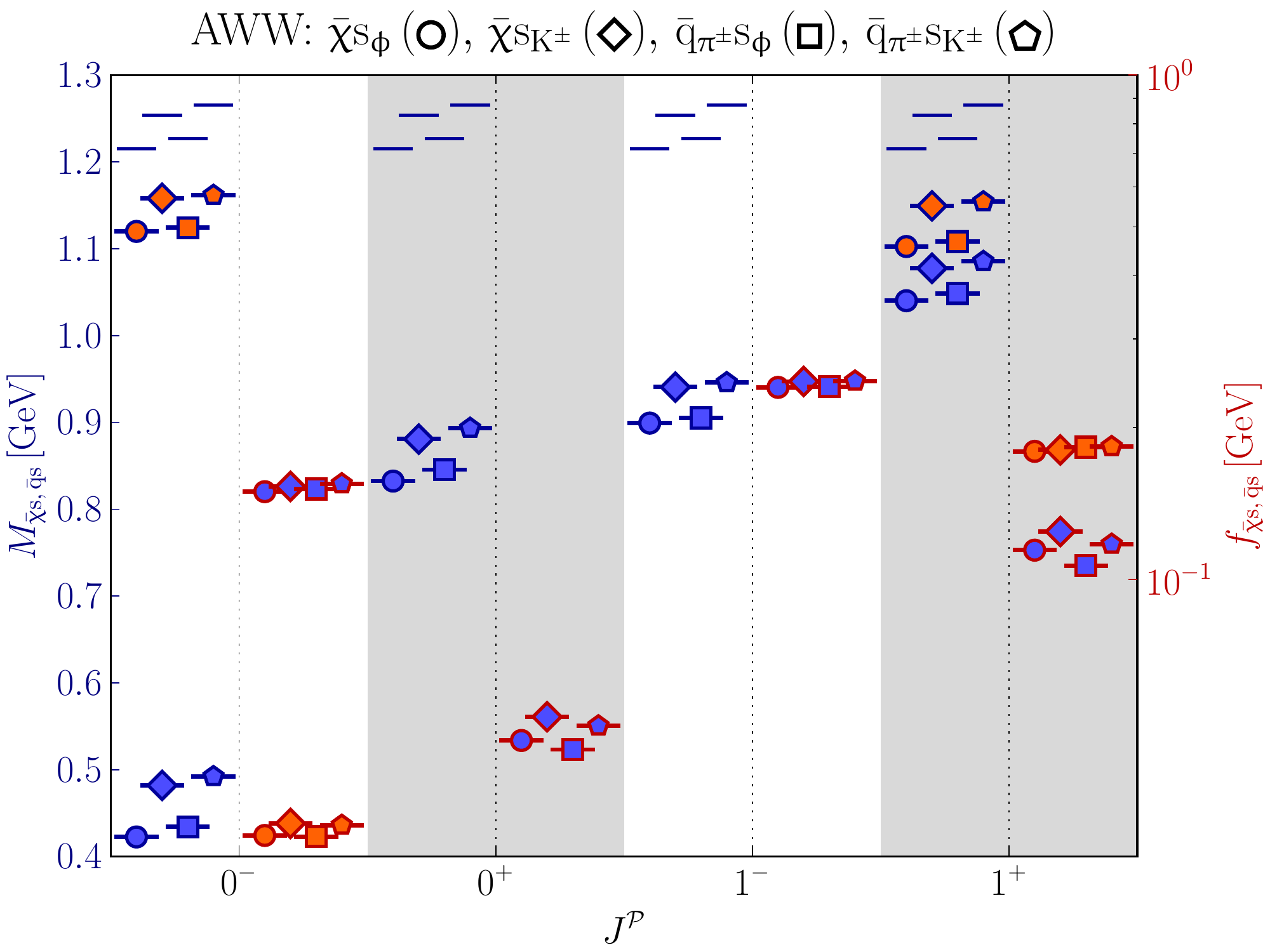}\caption{Strange mesons \cite[Figs.~14 and 5]{GRHKL}: Masses $M_{\bar\chi s,\bar qs}$ (left) and leptonic decay constants $f_{\bar\chi s,\bar qs}$~(right), from the ${\mathcal{G}}(k^2)$ couplings of Refs.~\cite{EI} (top) and \cite{EIa} (bottom). Note, \emph{e.g.}, an additional $0^+$ excitation in the~top.}\label{qs}\end{center}\end{figure}

\begin{figure}[h]\begin{center}\includegraphics[scale=.59784]{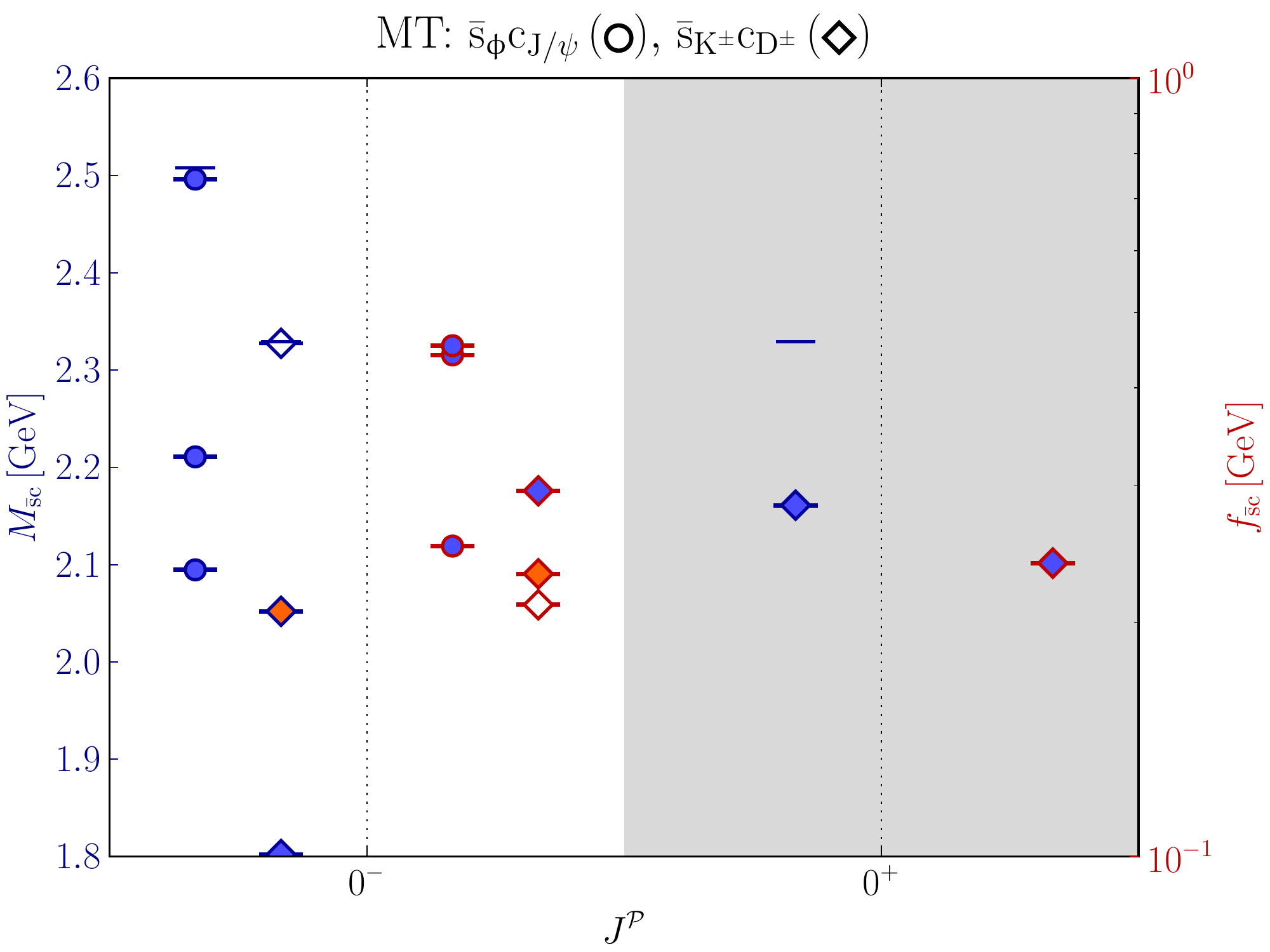} \includegraphics[scale=.59784]{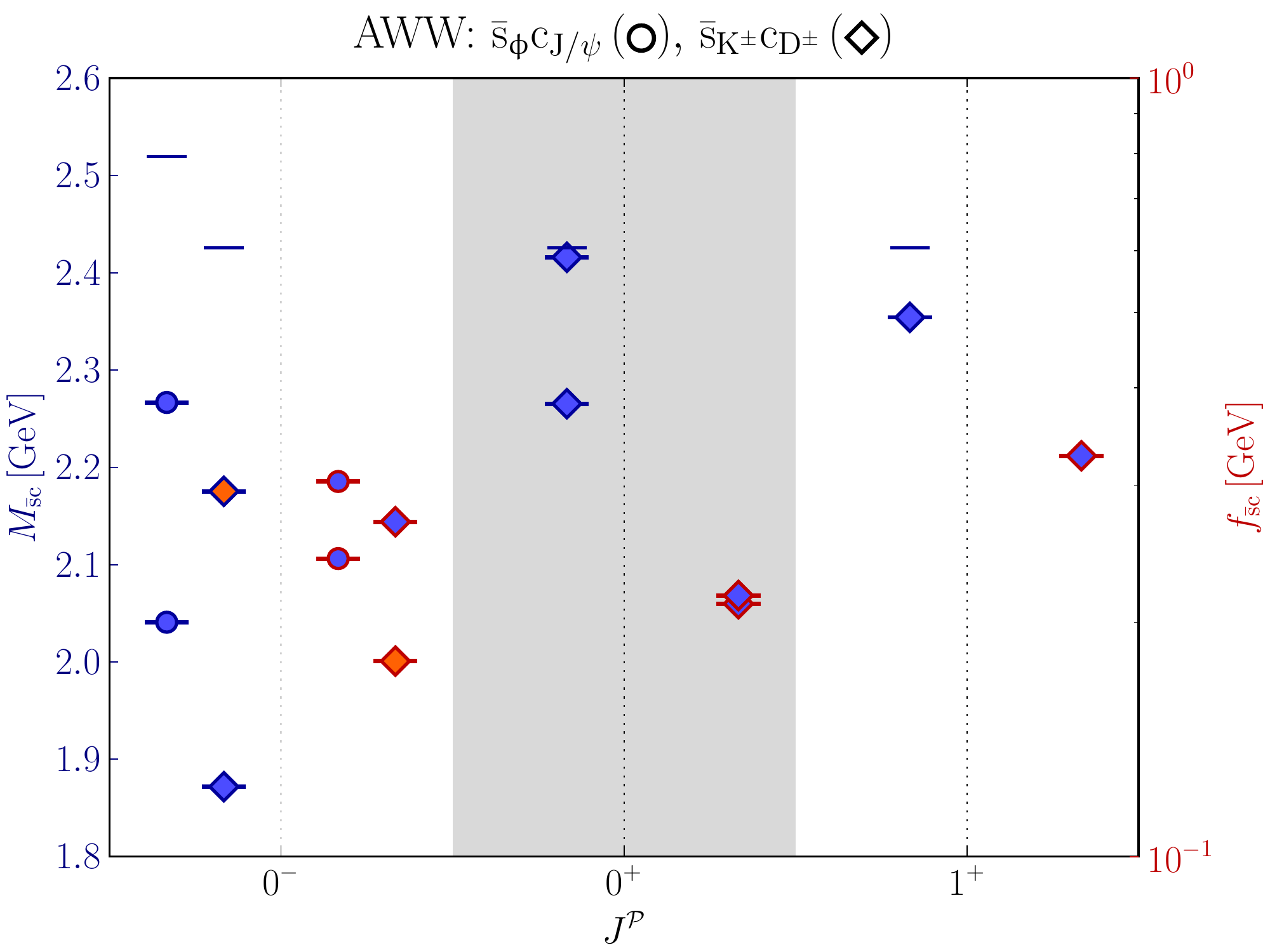}\caption{Charmed, strange mesons \cite[Figs.~17 and 8]{GRHKL}: Masses $M_{\bar sc}$ (left) and decay constants $f_{\bar sc}$ (right),~due to the ${\mathcal{G}}(k^2)$ ans\"atze of Refs.~\cite{EI} (top) and \cite{EIa} (bottom). Note the $1^+$ state in the bottom, not present in the~top.}\label{sc}\end{center}\end{figure}

\begin{figure}[h]\begin{center}\includegraphics[scale=.59784]{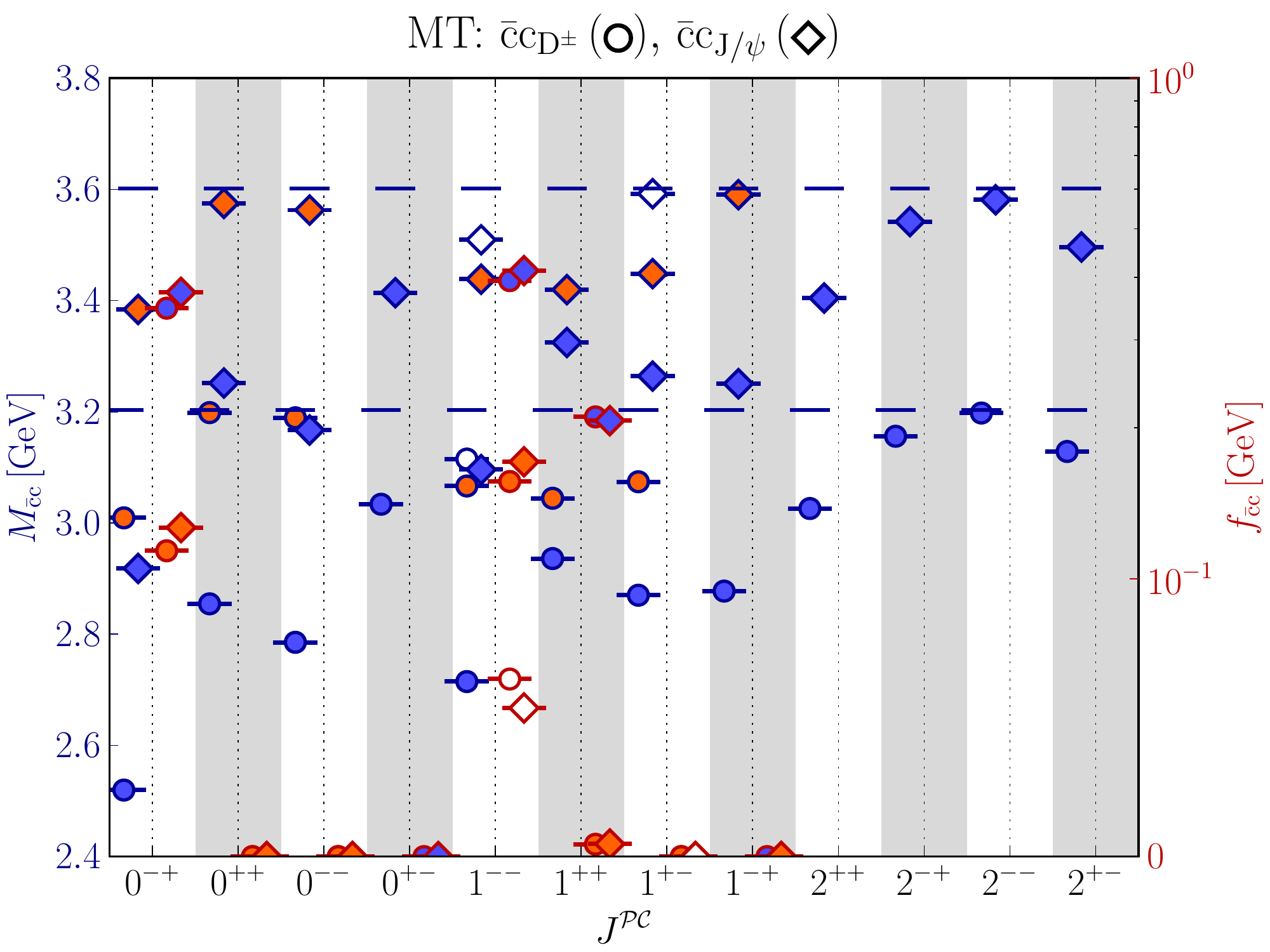} \includegraphics[scale=.59784]{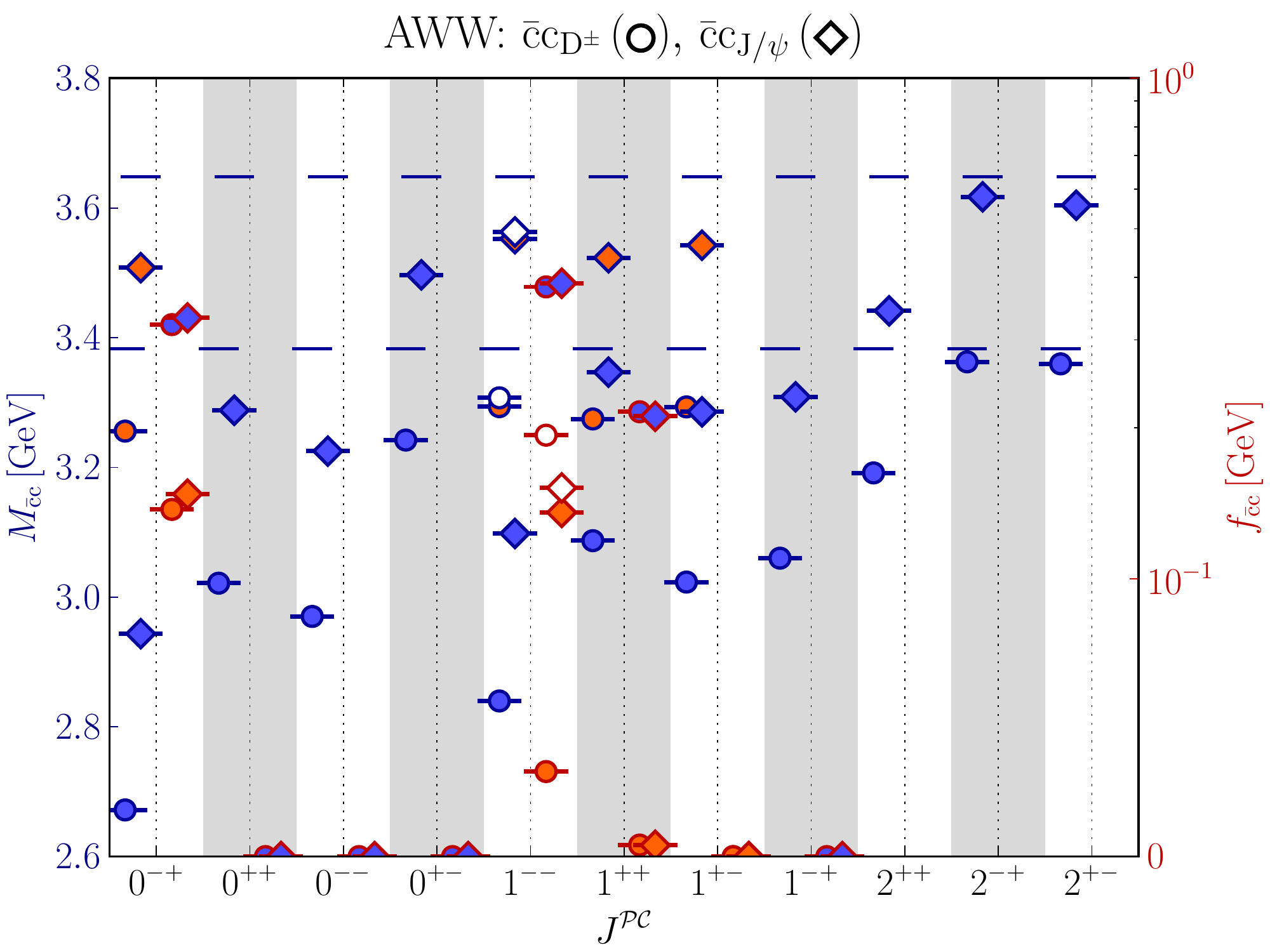}\caption{Charmonium \cite[Figs.~18 and 9]{GRHKL}: Masses $M_{\bar cc}$ (left) and decay constants $f_{\bar cc}$ (right), arising~from~the ${\mathcal{G}}(k^2)$ models of Refs.~\cite{EI} (top) and \cite{EIa} (bottom). The top results reveal many more ground and excited~states.}\label{cc}\end{center}\end{figure}

Covering the full mass range from a fictitious massless quark labelled $\chi$ up to the bottom quark, we organize our findings for masses and leptonic decay constants of meson ground states and lowest radial excitations \cite{GRHKL,WL@} in plots for spin, parity, and charge-conjugation parity combinations $J^{P\,(C)}$ as shown for the strange and charmed, strange mesons in Figs.~\ref{qs} and \ref{sc}, and for the charmonia~in~Fig.~\ref{cc}. 

\pagebreak

\begin{figure}[t]\begin{center}\includegraphics[scale=.59784]{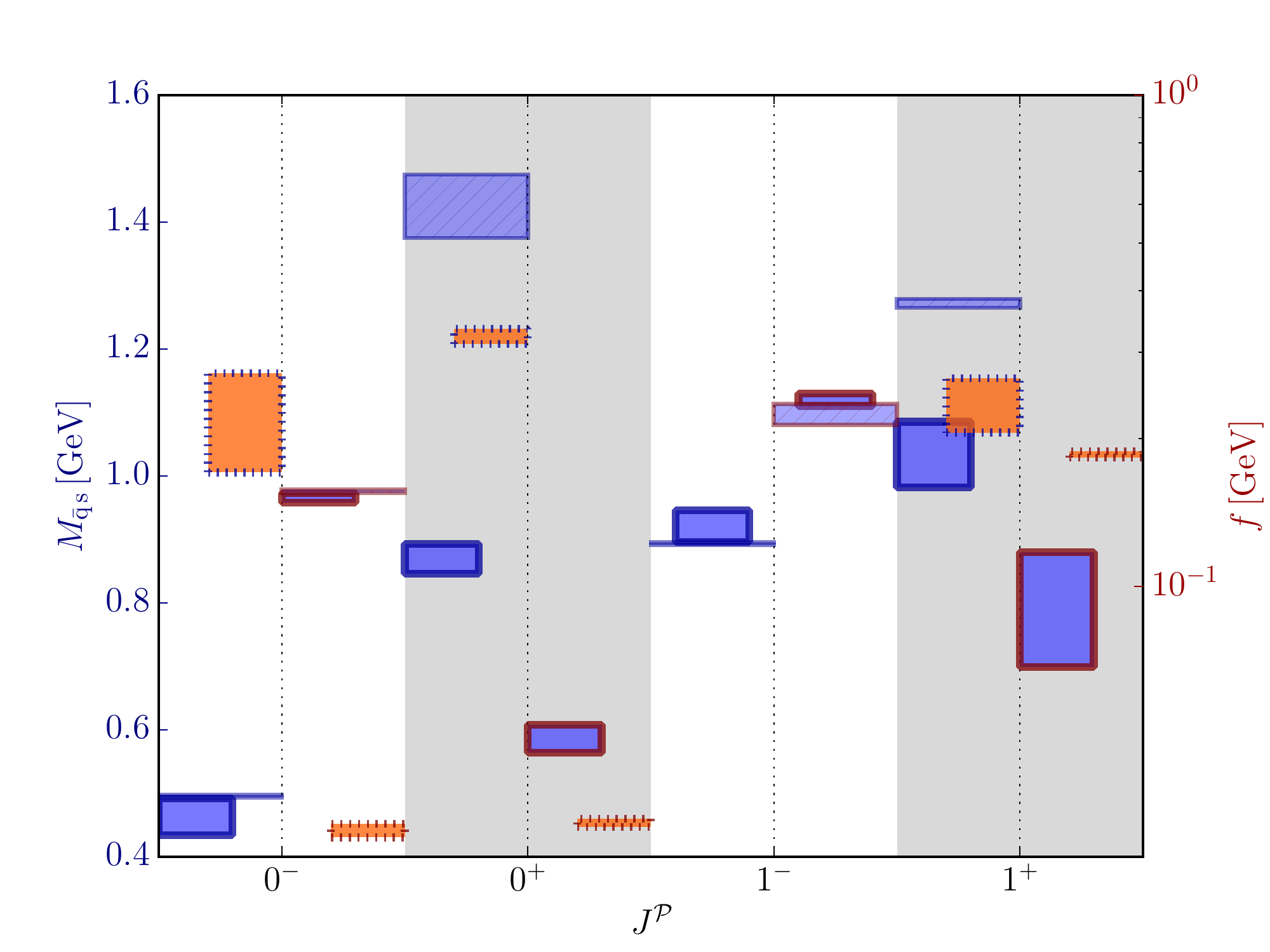} \includegraphics[scale=.59784]{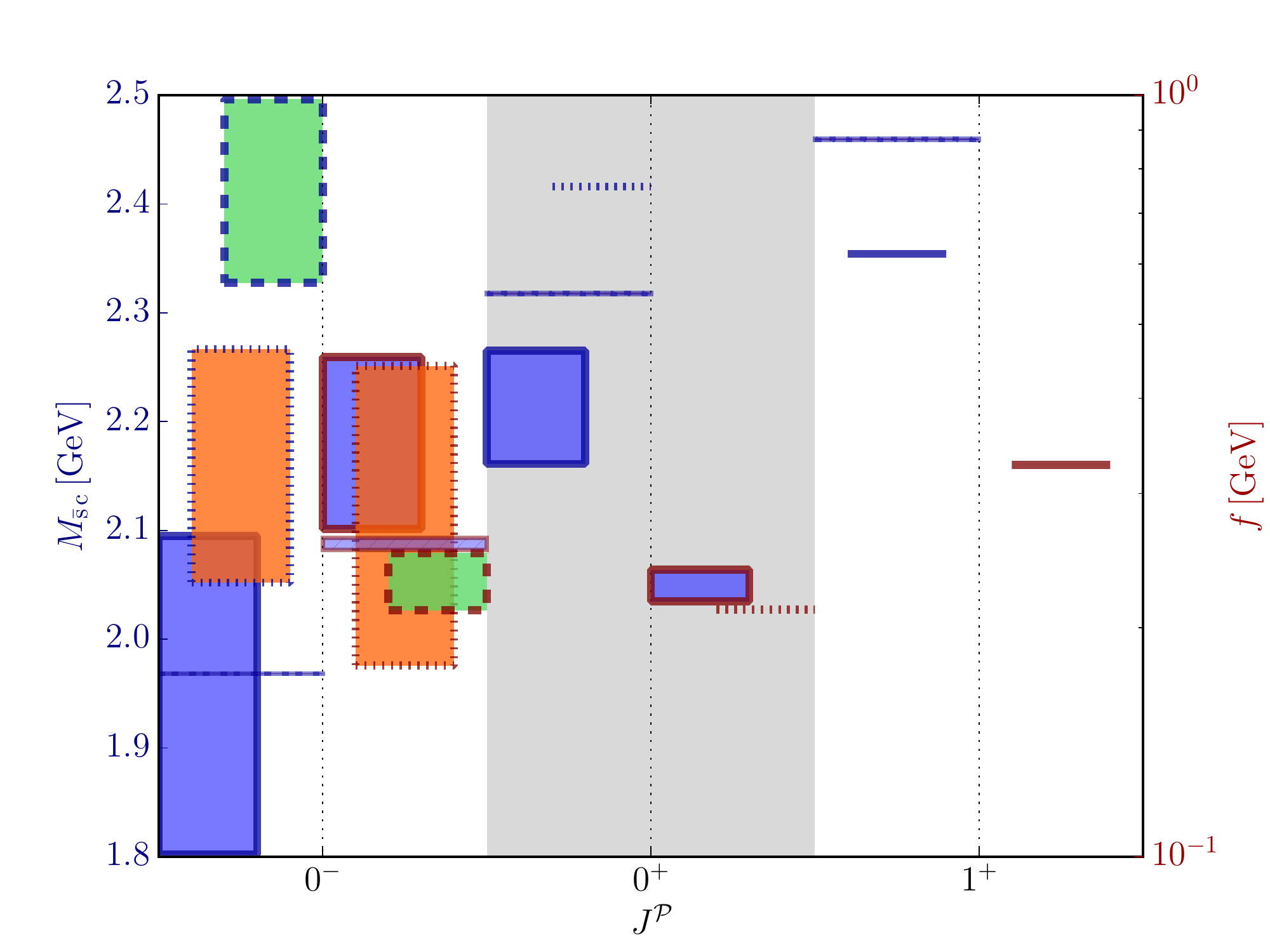}\caption{Mass spectra and leptonic decay constants of strange and charmed, strange mesons \cite[Figs.~24 and 25]{GRHKL}: combined results using both ${\mathcal{G}}(k^2)$ models (narrow boxes) confronted with experiment \cite{PDG} (wide~boxes).}\label{qse}\end{center}\end{figure}

\noindent Combining, for confrontation with experiment, the results of the adopted effective-coupling models (see, \emph{e.g.}, Fig.~\ref{qse}), we find an unexpectedly large model dependence, manifesting even in the~number of accessible states (\emph{i.e.}, not only in numerical meson properties): the interaction of Ref.~\cite{EI} tends~to provide more states. Our complete set of predictions can be found, also in form of tables, in Ref.~\cite{GRHKL}. 

\pagebreak

\begin{figure}[t]\begin{center}\includegraphics[scale=.59784]{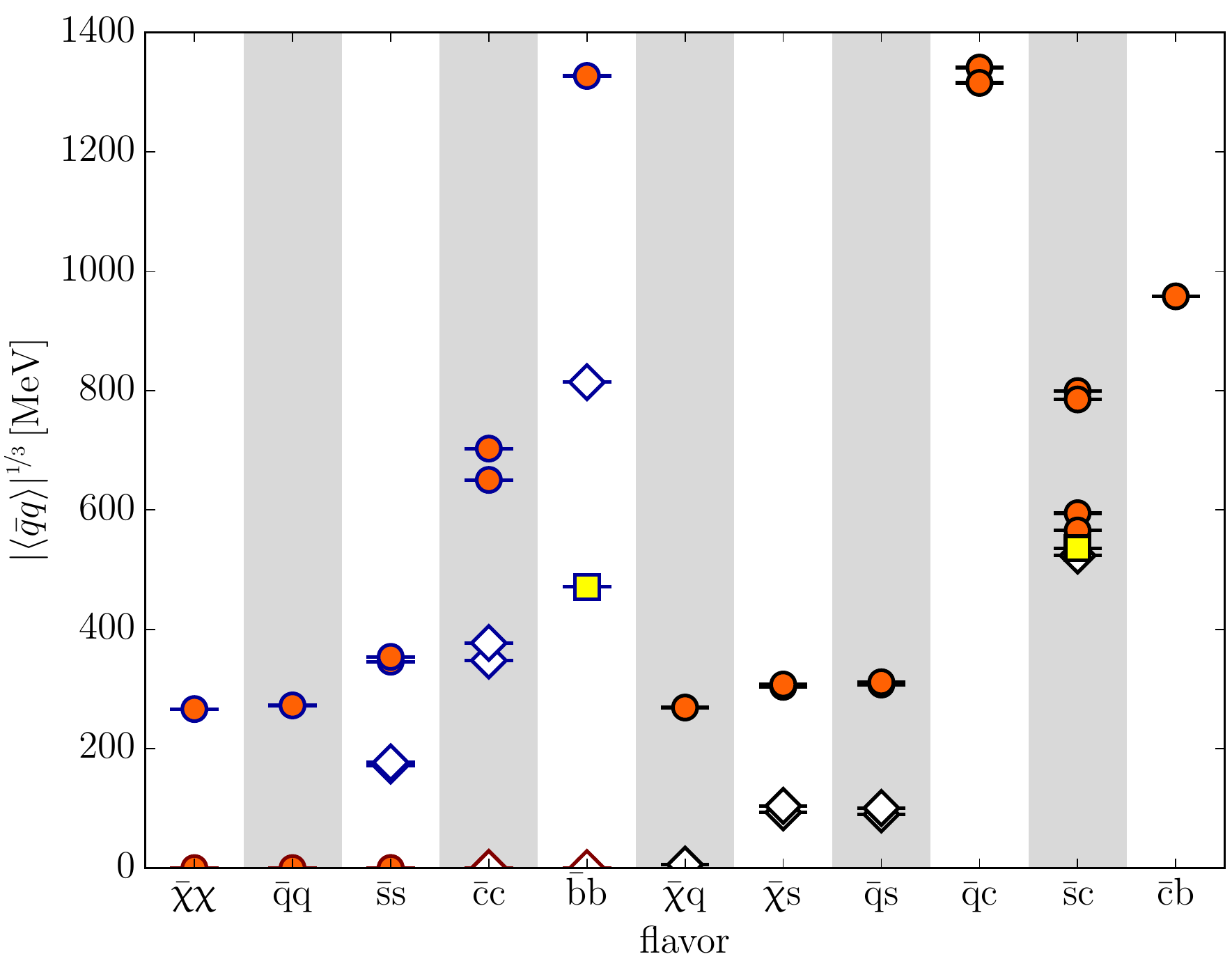} \includegraphics[scale=.59784]{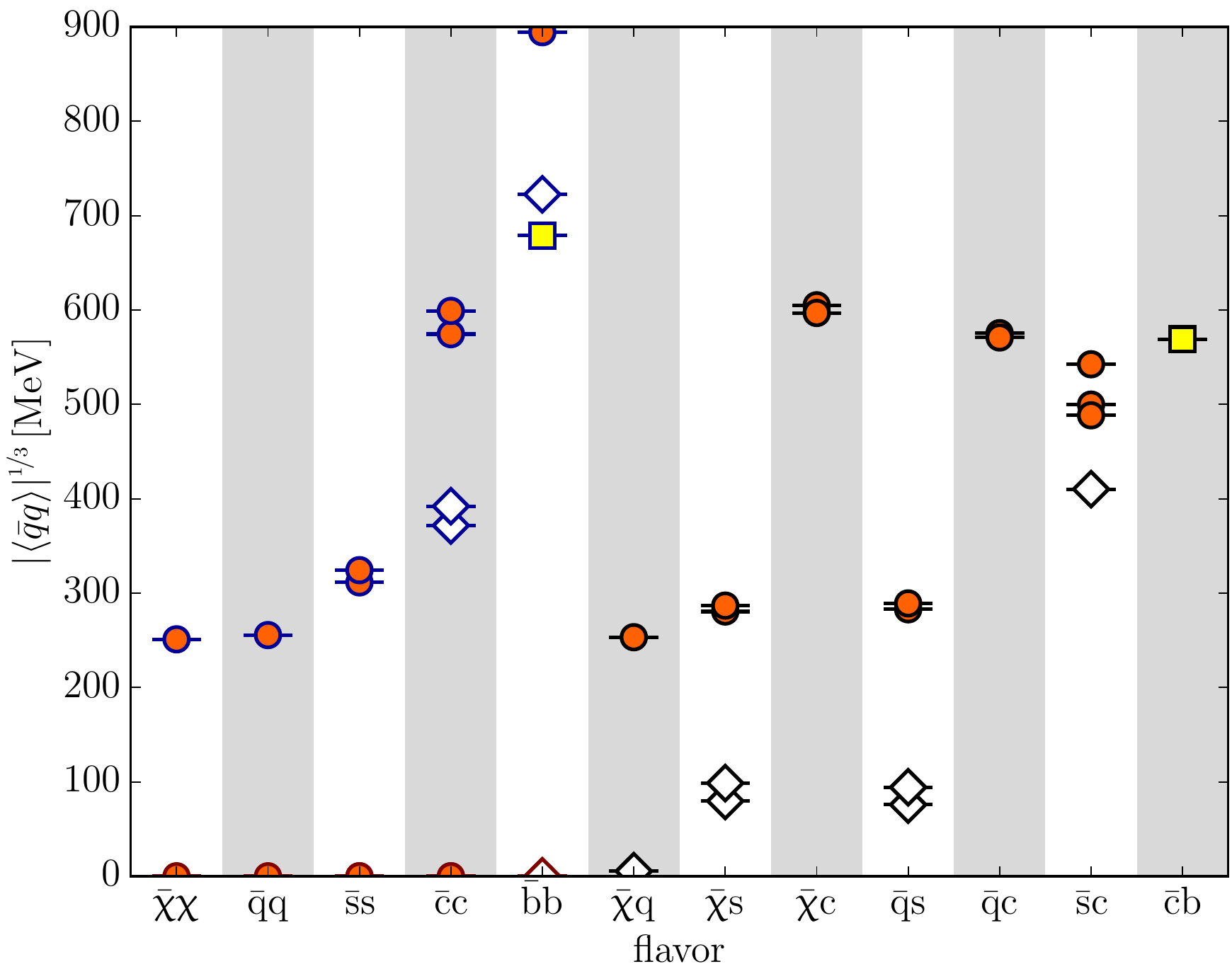}\caption{In-meson condensates \cite[Figs.~21 and 12]{GRHKL} for the ${\mathcal{G}}(k^2)$ ans\"atze of Refs.~\cite{EI} (top) and \cite{EIa} (bottom).}\label{ihc}\end{center}\end{figure}

Last, but not least, Fig.~\ref{ihc} presents the numerical values of the \emph{in-hadron condensate}, referred to as $\langle\bar q\,q\rangle$ \cite{MRT}. The latter quantity is defined, for some pseudoscalar meson $(\bar q\,q'),$ by~the~product~of~this state's leptonic decay constant $f_{\bar q\,q'}$ and projection onto an interpolating quark-bilinear~pseudoscalar operator $\bar q'\,\gamma_5\,q$ and satisfies a generalization of the good old Gell-Mann--Oakes--Renner relation \cite{MRT}.

\end{document}